\documentclass[]{egwr}

\newcommand{\eqref}[1] {(\ref{#1})}
\newcommand{\text}[1] {\mbox{#1}}

\begin{document}

%
\title{PVM-Distributed Implementation of the Radiance Code}

\author{F.\ R.\ Villatoro\footnote{Corresponding author}}
\institution{
Departamento de Lenguajes y Ciencias de la Computaci\'on\\
 E.T.S.\ Ingenieros\ Industriales\\
 Universidad de M\'alaga\\
 villa@lcc.uma.es
}
\author{A.\ J.\ Nebro and  J.\ E.\ Fern{\'a}ndez-Mart{\'\i}n}
\institution{
Departamento  de Lenguajes y Ciencias de la Computaci\'on\\
 E.T.S.\ Ingenieros\ en Inform\'atica\\
 Universidad de M\'alaga\\
antonio@lcc.uma.es
}

\maketitle


\begin{abstract}

The Parallel Virtual Machine (PVM) tool has been used for a
distributed implementation of Greg Ward's Radiance code. In order
to generate exactly the same primary rays with both the sequential
and the parallel codes, the quincunx sampling technique used in
Radiance for the reduction of the number of primary rays by
interpolation, must be left untouched in the parallel
implementation. The octree of local ambient values used in
Radiance for the indirect illumination has been shared among all
the processors. Both static and dynamic image partitioning
techniques which replicate the octree of the complete scene in all
the processors and have load-balancing, have been developed for
one frame rendering. Speedups larger than 7.5 have been achieved
in a network of 8 workstations. For animation sequences, a new
dynamic partitioning distribution technique with superlinear
speedups has also been developed.

\end{abstract}

\vspace{.5cm}


%
\section{Introduction}
\label{INTRODUCTION}

Radiance~\cite{Warduno, Warddos} is a global illumination package
based on hybrid approach that uses deterministic ray tracing, path
tracing and Monte Carlo techniques~\cite{Glassner}. It is a widely
used in architectural lighting system designs and visualization.
However, Radiance requires a considerable amount of computation
before an image can be generated, so its parallel implementation has
to be addressed.

The standard distribution of Radiance provides the possibility of
parallel execution of the rendering phase of a single image using a
memory sharing algorithm and the parallel/distributed execution
during the rendering of animations with many frames, where the
frames to be rendered are distributed between the processors each
one running the sequential code. However, better techniques have to
be developed.

The kernel for the path tracing of Radiance is a ray tracer and
the techniques developed for parallel ray tracing can be applied
straightforwardly. A ray tracer can be distributed by partitioning
of the image (pixel) space or the scene (object)
space~\cite{Jensen, Sung}. The techniques of image partitioning
yield better results but have the problem that the complete octree
of the scene has to be replicated in all the processors and so the
size of the image which can be rendered is limited by the amount
memory available in the computers.

Object partitioning techniques have the problem that the
ray/intersection process must test any ray with all the objects of
the scene, and this may require that the ray be traced by all the
processors~\cite{Kim}.

A hybrid image/space technique combining demand driven and data
parallel execution for the paralelization of Radiance using PVM has
been developed by Reinhard and Chalmers~\cite{Reinhard}. These
authors do not preserve the quincunx algorithm, do not distribute
the octree of local ambient values and do not obtain superlinear
speedups.

The purpose of this work is to distribute Radiance code in a
network of heterogeneous UNIX workstations using the Parallel
Virtual Machine (PVM) tool.
In this paper, several image partitioning techniques for the
distributed implementation of Radiance are studied. These
techniques are based on client/server architecture, where the
server receives the image/animation to be rendered, distributes
the work to be done between the clients, manage the load-balancing
techniques, collects statistical data on the rendering phase and
generates the output image.

In this paper several techniques for the image partitioning of the
image space are developed and tested using the parallel virtual
machine (PVM) tool~\cite{Geist}. In Section~\ref{RADIANCEBASICS},
the basic structure of Radiance is recalled and the quincunx
sampling algorithm is presented in more detail. In
Section~\ref{DISTRIBUTIONBASICS}, the basic techniques for the
distribution of a ray tracer are reviewed with emphasis on the
image partitioning techniques and, in
Section~\ref{DISTRIBUTEDRADIANCE}, it is applied to Radiance with
emphasis on the preservation of its acceleration techniques.
Section~\ref{STATICRADIANCE} is devoted a static image
partitioning technique with an estimator for the best partition,
and a new technique for load-balancing is developed. Dynamic image
partitioning techniques based on both scanbars and windows are
developed in Section~\ref{DYNAMICRADIANCE} including an
application to the rendering of animation sequences. The main
results and conclusions are presented in Sections~\ref{RESULTS}
and~\ref{CONCLUSIONS}, respectively.

%
\section{Radiance basics}
\label{RADIANCEBASICS}

Radiance~\cite{Warduno, Warddos} is a global illumination package
which solves the radiance or rendering equation, including
participating media, by path tracing with Russian roulette, and
selects new rays based on hybrid approach using both deterministic
ray tracing and Monte Carlo techniques with importance sampling
depending on the Bidirectional Reflectance Distribution Function
(BRDF)~\cite{Glassner}. Radiance uses a hybrid local/global
illumination model and allows for the specification of isotropic and
anisotropic BRDF.

Every object in a Radiance scene is a collection of modifiers,
e.g., surfaces and volumes which define the geometry, materials
which determine the illumination model parameters, textures,
patterns and mixtures of several modifiers. The materials in
Radiance include light sources, specular, Lambertian, isotropic and
anisotropic general BRDF material models, and participating media.

The Radiance ray tracing kernel includes three basic acceleration
techniques: the scene is stored in an octree in order to accelerate
the ray/object intersection process, the indirect illumination is
calculated during the path tracing and stored as local ambient
illumination in an octree, and the number of primary rays is
reduced using an adaptive partitioning of the image space including
interpolation of pixel colors by a quincunx sampling
technique~\cite{Warddos}.

The octree used by Radiance to subdivide the object space is
created by the program \verb$oconv$ of the Radiance package and is
independent of the renderer \verb$rpict$. In all the results and
comparisons made during this research, the time for the creation
and distribution of the octree is considered as a part of the total
computational time.

The octree that stores the local ambient values used for the
calculation of the indirect illumination in Radiance has exactly
the same structure as the one used for the scene. Each subcube of
this octree contains an unbounded list with the local ambient
illumination values calculated during the path tracing of the
objects inside it. To determine the indirect illumination at one
point in the scene, a hemisphere around this point is sampled using
adaptive supersampling, and, for each subdivision, the local
ambient values are collected. A measure of the expected error is
calculated and used for the adaptive supersampling process.


Radiance divides the image space (\verb&hres& $\times$ \verb&yres&
pixels) in samples (\verb&xstep& $\times$ \verb&ystep& pixels) and
uses a sampling technique, referred to as quincunx sampling, to
calculate the colors of the pixels in each sample~\cite{Warduno}. A
scanbar is a collection of \verb&ystep& consecutive scanlines each
one divided in samples of \verb&xstep& pixels. Note that both the
last scanbar and the last sample may have less number of pixels than
the rest. Two consecutive scanbar share a common scanline (except
for the first one), which is calculated only once. There are pixels
in a scanline for which no primary ray is traced and an
interpolation is made. The number of these pixels is determined by a
ray density vector which is updated during the quincunx sampling.

Each scanline is divided in samples of $\verb&xstep&+1$ pixels
sharing one pixel between consecutive samples. Odd and even
scanlines are sampled in a slightly different manner. Primary rays
are traced through the pixels of the extremes of the samples, and
other pixels inside the sample depending on the values of the ray
density vector; these values can change as a function of the
number of rays traced during the color filling of the sample. The
rest of the pixels in the sample are calculated by interpolation.

The color interpolation between pixels used in the horizontal or
vertical direction is based on a local density of rays (ray density
vector) and proceeds as follows. Given two extreme pixels in the
sample, if the difference between its colors exceeds a given
tolerance or the local density of rays for this sample is greater
than zero, a ray is traced by the center of the sample, and the
local density of this sample is changed to 1. Otherwise, the color
of the center pixel is calculated by interpolation and the local
density of this sample is change to 0. If there are pixels
remaining in the sample without color, the same procedure is
applied recursively to the two sub-samples defined by the center
pixel with a new local density of rays equal to a half of the
original one.

%
\section{Radiance distribution basics}
\label{DISTRIBUTIONBASICS}

Radiance provides the possibility of parallel execution of the
rendering phase of a single image using a memory sharing algorithm
(\verb$rpiece$). The shared memory stores the octree of local
ambient values used to accelerate the indirect illumination
calculation where these values are written as they are calculated
and are shared among all the processors. Radiance also provides a
simple mechanism for its parallel/distributed execution during the
rendering of animations with many frames, where the frames to be
rendered are distributed among the processors each one running the
sequential code.

The Parallel Virtual Machine (PVM) is a standard tool to distribute
a sequential code, e.g., Radiance, in a network of heterogeneous
UNIX workstations. This tool uses message passing as a basic
communication mechanism among a system of distributed tasks with
collaborate inside a virtual machine appearing as only one computer
with a common distributed memory. The main characteristics of
message passing in PVM are the following: the sender of a message
is asynchronous and can continue execution, the receiver can act
both synchronously, stop execution and wait for a message, or
asynchronously, check if there is any received message and continue
execution if not; it provides a multicasting service for the
sending of a message to all the processes; and, the message size is
not limited.

The heart of Radiance is a ray tracer and, therefore, can be
implemented in a distributed shared memory system by both
partitioning the image (pixel) space or the scene (object)
space~\cite{Jensen}.

In object space partitioning, the scene is divided amongst all the
processors. Since in a ray tracer, the more demanding task is the
ray/object intersection, object partitioning accelerates this
process because it reduces the number of objects in each
processor. Two main schemes can be used~\cite{Kim}: ray data flow
where rays are transferred through all the processors in order to
obtain the nearest intersection with the scene, and data object
flow where objects are moved through the processors and a virtual
memory stores all the objects and every processor uses a local
cache with the objects more frequently used. The success of both
schemes is based on the coherence of the scene and on the rays.

In image partitioning, the pixels of the image are distributed
amongst the processors and the scene is completely replicated. The
partitioning can be static or dynamic. The performance of static
partitioning depends on the technique of distribution of the
pixels among the processors. In order to avoid idle processors,
the parts of the image which are computationally more expensive to
render must be divided into smaller pieces or sent to more
powerful processors. Load balance can be improved during the first
distribution of the pixels amongst the processors based on
estimators of the computational cost or by redistribution of the
load when some processors become idle.

In dynamic partitioning of the pixels of the image, a pool of
pieces or samples of the image is built and the processors ask for
samples of the image as they become idle, render sequentially this
sample, and then request for another one. The size of these samples
of the image depends on the distribution technique used and the
nature of the algorithm used by Radiance to sample the image space.
Load balancing can also be applied when using dynamic partitioning.

Both static and dynamic partitioning techniques can obtain good
speedups, but with the drawback that the complete scene has to be
replicated in all the processors and the processor with smallest
memory capacity controls the size of the scene which can be
processed by these techniques.

\section{Distributed implementation of Radiance}
\label{DISTRIBUTEDRADIANCE}

Several image partitioning techniques for the distributed
implementation of Radiance are studied in the next sections. These
techniques are based on client/server architecture where the
server receives the image/animation to be rendered, distributes
the work to be done between the clients, manage the load-balancing
techniques, collect statistical data on the rendering phase, and
generate the output image.

Any parallel/distributed implementation must deal with the three
basic acceleration techniques used by Radiance: the quincunx
sampling technique, the octree representation of the scene, and the
octree of local ambient values used for the indirect illumination.

The most characteristic feature of Radiance is the quincunx
sampling algorithm and this must be preserved by any parallel
implementation in order to generate exactly the same number of
primary rays in both the sequential and parallel codes. This can be
reached by partitioning the image into scanbars and sharing the
common scanline between consecutive scanbars among the
corresponding processors.

In the image partitioning techniques developed in this paper, the
octree for the scene is generated by the server and completely
replicated among all the clients.

Each client has its own octree of ambient values and broadcasts to
the other clients all the new ambient values as they are
calculated. Every client has a parallel task to update the local
ambient values as they are received and without interference with
the render task. This guarantees that all the clients have a
consistent copy of the same octree of ambient values. Moreover,
this parallel updating of the ambient values may yield a reduction
of the total number of rays required for the indirect illumination,
since values in some parts of the scene are updated before the
sequential code does.

\section{Static partitioning of the image}
\label{STATICRADIANCE}

The image space is partitioned into scanbars which are distributed
among the processors. These scanbars share its last scanline and
attention had to be paid to avoid its recalculation by more than one
processor. However, this partition of the image cannot be uniform,
i.e., every processor cannot receive the same number of scanbars,
because the computational cost of each scanbar changes considerably
depending on the complexity of the scene visible inside each one. To
obtain better partitions, an estimator of the scene complexity,
i.e., the cost of each scanbar, must be developed.

In order to estimate the computational cost associated with a
scanbar, several primary rays selected randomly are traced through
the scanbar. In order to measure the cost for these rays, it is
possible to count both the total number of rays generated in the
rendering or the total number of ray/object intersections used. The
performance of both measures has been compared with the total
computational time for several images and both measures yield
significant information on the complexity of the scanbar.

After the estimation of the computational cost of every scanbar,
the scanbars are grouped in partitions of approximately the same
computational cost and distributed amongst the processors. If the
$i$-th partition has $n_i$ scanbars and its cost is $C_i(n_i)$,
then the static distribution technique tries to obtain
\[
 C_1(n_1) \approx C_2(n_2) \approx \cdots \approx C_p(n_p),
\]
where $p$ is the number of processors. If $c_j$ is the estimated
cost for the $j$-th scanbar, the above expression yields
\[
 \sum_{j=1}^{n_1} c_j \approx \sum_{j=n_1+1}^{n_1+n_2}
  \approx \cdots \approx \sum_{j=n_1 +\cdots+n_{p-1}}^{n_1 +\cdots+n_{p}},
\]
which can be solved iteratively by a procedure of equalization.

The clients, in the client/server model used for the static
distribution of the image among the processors, execute the
Radiance code sequentially on the partition of the image, i.e., the
set of scanbars, that they have received from the server. In order
to share the last scanline of its partition with another processor,
every client starts its execution by calculating this scanline and
transferring it to the corresponding processor, and then proceeds
with the rest of the partition as usual.

In order to obtain larger speedups, a load-balancing technique
controlled by the server has been implemented. The server receives
requests from the clients which become idle, collects information
on the work remaining to be done in every client, selects the
client with larger remaining work and, if this work is larger than
one scanbar, demands that this process transfer half of its
scanbars remaining to be rendered to the client who has become
idle, while this process continues its execution. The last
scanline shared among both processors is rendered by the client
recently receiving the new partition which automatically resends
it to the first client. It may be possible that the first client
receives another request to transfer part of its work to a third
client before it receives the common scanline; in that case, the
second client is notified to send the previously common scaline to
this new client.

\section{Dynamic partitioning of the image}
\label{DYNAMICRADIANCE}

In dynamic partitioning, the pixels are redistributed among the
processors so there is no idle one. In a client/server architecture,
the server holds a pool of pieces of the image and distributes this
work to the clients as they make requests when they become idle. Two
solutions have been developed depending of the pieces used for the
partition of the image, i.e.,  scanbars that preserve exactly the
quincunx algorithm, or windows that mantain the quincunx algorithm
only on windows. Also a dynamic partitioning technique for animation
sequences has been developed.

\subsection{Scanbar partition of the image}

In this implementation, every client renders a scanbar locally. In
order to share the last scanline among the processors, each client
calculates its first scanline and transfers it to another
processor, and receives the last scanline from another client.
Automatic load-balancing is obtained.

The server receives the requests from and assigns scanbars to the
clients. With $p$ clients, the first $p$ scanbars that the server
assigns are not consecutive. Later, when sending a new scanbar, the
server checks if it can assign a scanbar consecutive to the scanbar
the client has finished  rendering in order that the last scanline
of the previous scanbar is shared locally, without communication
through the network. If no consecutive scanbar can be assigned,
another one is selected, and the client currently processing the
preceding scanbar is advised on the processor holding its
consecutive scanbar.

Every client renders a scanbar by the following steps: 1) It
calculates the first scanline; at any moment, when it knows which
processor is processing the scanbar sharing this scanline, it sends
it automatically; this first scanline is sent incrementally as
every \verb$xstep$ pixels are calculated. 2) It renders the last
scanline if the scanbar is the first one of the image. 3) It
applies quincunx algorithm to fill the rest of the scanlines of the
scanbar by interpolation with the quincunx sampling algorithm; this
process is done incrementally as the last scanline is received, if
required. This procedure avoids that shared scanlines of high cost
in a processor delayed excessively the processor sharing this
scanline.

\subsection{Window partition of the image}

In all the distribution solutions for the Radiance code presented
above the quincunx sampling algorithm for the selection of the
primary rays was left as it was. Another possibility is to divide
the image into uniformly sized windows. These windows may be
distributed among the processors and a client/server architecture
may be adopted. The server holds a pool of windows with are sent
when a processor becomes idle. No sharing of scanlines is
required. The client renders the window as the sequential code
does, adjusting appropriately the camera parameters. The only
information share among processors is the values of indirect
illumination that can be calculated in the other machines.

With this partition, the number of generated primary rays exceeds
the sequential code and grows with the number of windows. The
reasons for the increase in the number of rays are that both the
number of total scanlines, although they do have shorter length,
and the number of primary rays for the image sampling are
increased; moreover, an increase in the number of primary rays
increases the ray density vector values and further rays are
traced. Note that for a few exceptional windows, the number of
primary rays traced is even smaller.

The algorithm that the server uses to select the next window to be
transferred to an idle process can affect the speedup. The best
strategy for the selection of the new window is to send first the
windows with higher cost and then those of lower, but a priori is
impossible to determine this order exactly. Three strategies:
backward, forward and random selection have been implemented. An
experimental study indicates that the differences in the speedups
achieved with these three strategies are small even when the scenes
used in the tests have clear differences in computational cost
between the first or the last windows and the rest. Of course, the
difference comes from the rendering of the last windows, where some
processors can become idle without further work to be done. As the
number of windows is increased these differences decrease.

\subsection{Dynamic partitioning of an animation sequence}
\label{ANIMATION}

Usually Radiance is employed to obtain an animation sequence, e.g.,
a virtual walk through a building. This animation is usually
rendered in a frame by frame manner. This requires that the global
illumination information be recalculated for every frame.

The dynamic partitioning technique based on windows presented in
the previous section, where all the frames of the animation are
divided into windows, has been used in this paper. When an idle
client asks the server for a new window, it selects a window of the
present frame (if any remains) or one of the next frame
(otherwise). Both client and server must know the actual frame
being rendered.

\section{Presentation of results}
\label{RESULTS}

The techniques developed in this paper for the distributed
implementation of the Radiance code have been tested in an ATM
network of 8 Ultra Sparc-2 workstations. It is important to note
that the network used for the tests presented in this section has
a low load and the processors run only the present application;
therefore, the communications were very fast. Shareware versions
of the Radiance 3.0 code and the PVM 3.3.1 tool have been used.
All the scenes used in the tests presented in this section can be
obtained from the standard distribution of Radiance.
Table~\ref{tab:escenas} present its main characteristics.

\begin{table}
\caption{Scenes used in the presentation of results from the standard
distribution of Radiance. The CPU time is in hours on a Ultra
Sparc-2 workstation.}
\vspace{.5cm}

\begin{center}
\begin{tabular}{||c|cccc||}
 \hline
 Scene & Filename  & Resolution & Rays traced & Time \\ \hline
 1 & \verb&Mmack Leftbalc& & 2000$\times$1352 & 97,128,720 & 25.598 \\
 2 & \verb&Conference room& & 1000$\times$676 & 27,146,574 & 0.914 \\
 \hline
\end{tabular}
\end{center}
 \label{tab:escenas}
\end{table}

In all the algorithms presented in this paper, a consistent copy of
the octree of local ambient values was distributed among all the
processors. Table~\ref{tab:broadcastambient} shows the importance of
this distribution in order to reduce both the number of rays and the
real computational time. Although Table~\ref{tab:broadcastambient}
shows the results for the dynamic partitioning technique based on
windows, similar results have been obtained for the rest of the
techniques presented in this paper. This table shows that the number
of rays is reduced about 55\%--65\% and the computational cost is
reduced about 45\%--55\% thanks to the sharing of the local ambient
values.

\begin{table}

\caption{Number of rays (in million) and real computational time
(in hours) for scene~1 with the dynamic partitioning based on
windows with 4 clients and several window numbers both with and
without the replication (broadcast) of the indirect illumination
local ambient values.}
\vspace{.5cm}

\begin{center}
\begin{tabular}{||c|c|ccc||}
 \hline
         &  Windows  & 64 & 1024 & 4096 \\ \hline
 Without Broadcast &  Number Rays & 146.96 & 174.40 & 177.72  \\
         &  Real Time   & 13.14 & 11.88 & 12.30 \\ \hline
 With Broadcast &  Number Rays & 97.961 & 96.956 & 97.969  \\
         &  Real Time   & 6.535 & 6.407 & 6.706 \\
 \hline
\end{tabular}
\end{center}
        \label{tab:broadcastambient}
\end{table}

In order to assess the behaviour of the estimator of the scanbar
cost used in the static partitioning technique, comparisons between
the partition based on the estimators and the optimum partition,
calculated after the completion of the rendering of the image, have
been developed. The corresponding results indicate that the
performance of the estimator for a scanbar increases as the cost for
this scanbar increases; when a large number of estimation rays are
generated, there is no guarantee that the estimation improves
because there are estimation rays which are not traced in the
standard rendering due to the quincunx sampling algorithm, and this
can degrade the estimation. These results also indicate that there
is a clear correspondence between the estimated complexity of the
scanbar of the image and the visually apparent complexity of these
scanbars.

For an image as scene 2, whose cost is nearly a smooth function of
the number of scanbars, Table~\ref{tab:static} shows that for these
scanbars of low complexity the estimator with 5 rays behaves better
than the estimator with 20 rays and that the estimator based on the
ray/object intersection is better than the one based on the total
number of rays generated; moreover, as the number of processors
increases these differences also increases. For an image as scene 1,
whose cost function has 7 peaks of high cost and a series of
plateaus of low cost, the performance of the estimator with 5 rays
is worse than that with 20 rays because of the large complexity of
this scene, but, in any case, the resulting partitions are not very
good, specially because of the high cost of the first scanbar. In
summary, the estimator with 20 rays is worse than the one with 5
rays (counting the number of rays or ray/object intersections); for
a low degree of partitioning, the estimator works better and the
ray/object intersection measure of cost is better than that of
generated rays.

\begin{table}

\caption{Speedup for scene 2 for the static image partitioning
without load-balancing using the following partitions of the
image: uniform, optimum, estimated with 5 and with 20 rays based
on the number of total rays (Estim.5a, Estim.20a) and estimated
with 5 and 20 rays based on the number of ray/object intersections
(Estim.5b, Estim.20b).} \vspace{.5cm}

\begin{center}
\begin{tabular}{||c|ccc||}
 \hline
 Processors & 2 & 4 & 8 \\ \hline
 Uniform    & 1.30 & 2.45 & 4.66 \\
 Estim.5a   & 1.99 & 3.96 & 6.01 \\
 Estim.5b   & 2.08 & 3.96 & 7.25 \\
 Estim.20a  & 2.09 & 3.60 & 6.22 \\
 Estim.20b  & 2.02 & 3.89 & 6.98 \\
 Optimum    & 2.07 & 4.01 & 7.37 \\
 \hline
\end{tabular}
\end{center}
        \label{tab:static}
\end{table}

Table~\ref{tab:staticloadbalance} shows the speedup obtained with
the static image partitioning with load-balancing. For 2 and 4
processors, slightly superlinear speedups are obtained because of
the distribution of the octree of local ambient values; this
result was to be expected. This table also shows that as the
initial partition is improved, higher speedups are obtained, since
the communication overload during the last phase of the rendering
is reduced.

\begin{table}

\caption{Number of rays (in million) and speedup for scenes 1 and
2 for the static image partitioning with load-balancing using as
initial partition of the image the partition with estimation of
ray/object intersection with 5 rays. The cost of this estimation
is included in the speedup calculation.} \vspace{.5cm}

\begin{center}
\begin{tabular}{||c|c|cccc||}
 \hline
         & Processors  & 1 & 2 & 4 & 8 \\ \hline
 Scene 1 & Number Rays & 97.1 & 96.1 & 95.1 & 96.1 \\
         & Speed-up    & 1 & 2.05 & 4.02 & 7.26 \\ \hline
 Scene 2 & Number Rays & 27.2 & 27.1 & 27.2 & 27.2 \\
         & Speed-up    & 1 & 2.12 & 4.00 & 7.43 \\
 \hline
\end{tabular}
\end{center}
        \label{tab:staticloadbalance}
\end{table}

Table~\ref{tab:dynamic} shows the number of rays and speedup for
the dynamic image partitioning based on scanbars. This table
indicates that the preservation of the quincunx algorithm by means
of this solution, yields a number of rays nearly constant and equal
to that of the sequential code; it also shows very good speedups of
about 7.5 for 8 processors.

\begin{table}
\caption{Number of rays (in million) and speedup for scenes 1 and 2
for the dynamic image partitioning based on scanbars.}
\vspace{.5cm}

\begin{center}
\begin{tabular}{||c|c|cccc||}
 \hline
         & Processors  & 1 & 2 & 4 & 8 \\ \hline
 Scene 1 & Number Rays & 97.1 & 97.1 & 97.1 & 97.1 \\
         & Speed-up    & 1 & 1.70 & 3.84 & 7.46 \\ \hline
 Scene 2 & Number Rays & 27.2 & 27.2 & 27.2 & 27.2 \\
         & Speed-up    & 1 & 1.76 & 3.96 & 7.62 \\
 \hline
\end{tabular}
\end{center}
 \label{tab:dynamic}
\end{table}

The number of rays and speedup for the dynamic image partitioning
based on windows are shown in Table~\ref{tab:dynamicwindows}. This
table shows that superlinear speedup has been obtained for some
partitions of scene 2 because of the reduction of the number of
indirect illumination rays due to the sharing of the octree of
ambient values. For example, for scene 2, the sequential code need
27.23 million of rays, but with 64 windows and 2 clients only
27.06 million, and with 4 clients only 27.02 million. However, the
number of primary rays increases as the number of windows
increases. For example, scene 2 with 1024 windows needs 374,424
rays more than with 64 windows, while scene 1 with 4096 windows
exceeds in 505,000 rays that for 256 windows.

Table~\ref{tab:dynamicwindows} also shows that the speedup with 8
clients for scene 2 is low due to the low cost of every window.
For the more complex scene 1, this is not the case and a better
performance is obtained with a larger number of windows. This
behaviour is due to the tradeoff between computational and
communication costs because for high computational cost, the
communication cost appears to be less significant.

\begin{table}

\caption{Number of rays (in million) and speedup for scenes 1 and 2
for the dynamic image partitioning based on windows as a function
of the number of windows and clients.}
\vspace{.5cm}

\begin{center}
\begin{tabular}{||c|ccccc||cccc||}
  \hline
  & \multicolumn{5}{c||} {scene 2} & \multicolumn{4}{c||} {scene 1} \\  \hline
 Windows & 4 & 16 & 64 & 256 & 1024 &  64 & 256 & 1024 & 4096 \\ \hline
 2 clients & 2.02 & 2.02 & 2.05 & 2.02 & 1.99 &  1.89 & 1.91 & 1.99 & 1.82 \\
 4 clients & 2.44 & 3.81 & 4.03 & 3.96 & 3.54 &  3.92 & 3.98 & 3.99 & 3.82 \\
 8 clients & 2.64 & 6.44 & 6.87 & 7.14 & 6.48 &  7.36 & 7.79 & 7.87 & 7.76 \\
 \hline
\end{tabular}
\end{center}
 \label{tab:dynamicwindows}
\end{table}

The technique of dynamic partitioning based on windows of
animation sequences has been tested with two ``disconnected"
animations, where there is a large change on the camera position
from frame to frame, which is the worst case found in an
animation. A sequence based on the \verb$Mmack$ scene with 12
frames and using 8 clients requires 15.42 hours of CPU time if all
the frames are calculated independently using a dynamic
partitioning with 25 windows and only 5.98 hours using the dynamic
partitioning of the complete animation. Another animation, based
on the \verb$Townhouse$ scene of 17 frames, needs 109.45 CPU hours
with the sequential code, 15.4 hours partitioning frame by frame
and only 5.80 hours with the partitioning of the complete
animation; this corresponds to an extremely superlinear speedup of
18.3. The reason for this superlinear behaviour is twofold:
firstly, the use of the same octree for the scene throughout the
execution saves the time required to load and initialize the
octree for each frame and, secondly, the octree of ambient values
is calculated and shared among the processors.

\section{Conclusions}
\label{CONCLUSIONS}

This paper gives a general description and performance results for a
distributed version of Greg Ward's Radiance code by means of the
Parallel Virtual Machine (PVM) tool. Their implementation uses the
same primary rays as the sequential code, preserving the quincunx
sampling algorithm, and explores several strategies for static and
dynamic load-balancing. Client/server distributed implementations
for both single images and animation sequences have been developed.
In all these techniques, a copy of the complete octree of the scene
is replicated among all the processors therefore the smallest memory
size of any processor limits the size of the largest image which can
be rendered.

The results presented in this paper show that, trying to preserve
the quincunx sampling algorithm by using the scanbars as
distribution quantum, has the problem that a scanbar is a quantum
which for some images is very complex and of high cost, hence a
window partitioning with smaller windows than the scanbars yields
better results in practice.

In order to render large images, an object partitioning technique
which distributes the octree between all the processors has to be
developed. But this technique may have load imbalance problems with
intense communications. Further research on object partitioning of
Radiance is required. Also, the development of better estimators
which detect parts of the image with high/low cost is also important
although difficult in practice.






\begin{thebibliography}{99}

\newcommand{\paperref}[6] { {#1}. {#2}. {\textit {#3}},
  {#4}:{#5}, {#6}.}

\newcommand{\bookref}[4] { {#1}. {\textit {#2}}. {#3}, {#4}.}

\newcommand{\bookedit}[7] { {#1}. #2.
   In {#4}, eds., {\textit {#3}},  pp. {#6}, {#5}, {#7}.}

\newcommand{\progref}[4] { {#1}. { {#2}}. {#3}, {#4}.}


\bibitem{Warduno} \progref {G. J. Ward}
  {The RADIANCE 3.0 Synthetic Imaging System}
  {Lighting Systems  Research Group, Lawrence Berkeley Laboratory}
  {1996}

\bibitem{Warddos} \paperref{G. J. Ward}
  {The Radiance Lighting Simulation and Rendering System}
  {Computer Graphics}
  {28}{459--472}{1996}

\bibitem{Glassner} \bookref {A. S. Glassner}
  {Principles of Digital Image Synthesis}
  {Morgan-Kauffman Publishers}
  {1995}

\bibitem{Jensen} \progref {D. W. Jensen and D. A. Reed}
  {A Performance Analysis Exemplar: Parallel Ray Tracing}
  {Department of Computer Science, University of Illinois}
  {1992}

\bibitem{Sung} \paperref{K. Sung, J. L. J. Shiuan and A. L. Ananda}
  {Ray Tracing in a Distributed Environment}
  {Computer and Graphics}
  {20}{41--49}{1996}

\bibitem{Kim} \paperref{H.-J. Kim and Ch.-M. Kyung}
  {A New Parallel Ray-Tracing System Based on Object Decomposition}
  {The Visual Computer}
  {5}{244--253}{1996}

\bibitem{Reinhard} \bookedit {E. Reinhard and A. Chalmers}
  {Message Handling in Parallel Radiance}
  {Recent Advances in Parallel Virtual
     Machine and Message Passing Interface. Lecture Notes in Computer Science, Vol.  3241}
  {D. Kranzlmüller, P. Kacsuk, and J. Dongarra}
   { Springer-Verlag}
   {486--493} {1997}

\bibitem{Geist} \progref {A. Geist, A. Beguelin, J. Dongarra, W. Jiang,
  R. Manchek and V. Sunderam}
  {PVM 3 User's Guide and Reference Manual}
  {Oak Ridge National Laboratory, Oak Ridge, Tennessee}
  {1993}


\end{thebibliography}
\end{document}